\documentclass[aps,prb,twocolumn,superscriptaddress,showpacs]{revtex4}
\usepackage{graphicx}
\usepackage{calc}
\usepackage{bm}

\begin{document}

\title{Interference effects in transport across a single incompressible strip at the edge of the fractional quantum Hall system}

\author{E.V.~Deviatov}
\email[Corresponding author. E-mail:~]{dev@issp.ac.ru}
 \affiliation{Institute of Solid State
Physics RAS, Chernogolovka, Moscow District, 142432, Russia}

\author{B.~Marquardt} 
\affiliation{Laboratorium f\"ur Festk\"orperphysik, Universit\"at
Duisburg-Essen, Lotharstr. 1, D-47048, Duisburg, Germany}

\author{A.~Lorke}
\affiliation{Laboratorium f\"ur Festk\"orperphysik, Universit\"at
Duisburg-Essen, Lotharstr. 1, D-47048, Duisburg, Germany}

\author{G.~Biasiol}
\affiliation{Laboratorio Nazionale TASC INFM-CNR, AREA Science Park, I-34012 Trieste, Italy}

\author{L.~Sorba}
\affiliation{Laboratorio Nazionale TASC INFM-CNR, AREA Science Park, I-34012 Trieste, Italy}
\affiliation{NEST INFM-CNR and Scuola Normale Superiore, I-56126 Pisa, Italy}

\date{\today}

\begin{abstract}
We experimentally investigate interference effects in transport across a single incompressible strip at the edge of the quantum Hall system by using a Fabry-Perot type interferometer. We find the interference oscillations in transport across the incompressible strips with  local filling factors $\nu_c=1, 4/3, 2/3$ even at high imbalances, exceeding the spectral gaps. In contrast, there is no sign of the interference in transport across the principal Laughlin $\nu_c=1/3$ incompressible strip. This indicates, that even at fractional $\nu_c$, the interference effects are caused by "normal" electrons. The oscillation's period is determined by the effective interferometer area, which is sensitive to the filling factors because of screening effects. 
\end{abstract}

\pacs{73.40.Qv  71.30.+h}

\maketitle

\section{Introduction}

Interference phenomena in semiconductors have recently attracted considerable interest~\cite{heiblum,goldman,litvin,glattli,fabryperot}. A special and very exciting topic is the interference in the fractional quantum Hall  (QH) regime. This regime is characterized by a strong electron-electron interaction, which leads to the formation of a new ground state with unique physical properties~\cite{laughlin,haldane,obzor}. Interference oscillations were used to probe the fractional charge of the excitations and the fractional statistics in the FQHE regime~\cite{goldman}. 

In most experiments, the interference scheme was realized by using the edge state transport~\cite{buttiker}. Current-carrying edge states are arising at the sample edge at the intersections of the Fermi level and distinct Landau levels. Split-gates are used to  bring two different edge states into contact in two regions, called as quantum point contacts (QPC)~\cite{heiblum,goldman,litvin,glattli}. While moving along the edge state, the possible electron's path is divided into two at the first QPC, which are further reconnected  at the second QPC. Sweeping of the magnetic field allows to vary the phase difference between these two paths by penetrating a flux quantum through the effective interferometer area.  It produces the  interference oscillations of the current through the device~\cite{heiblum,goldman,litvin,glattli}.

The real sample edge potential is smooth, which  gives rise to the compressible-incompressible strips formation~\cite{shklovsky}. Landau levels are pinned to the Fermi level in some regions (compressible strips), while the local filling factor $\nu_c$ is constant in others (incompressible ones). This picture is especially applicable to the electrostatic  potential profile in the QPC region~\cite{shklovsky}.   The presence of the compressible regions does not destroy the interference in the integer QH regime~\cite{fabryperot}, because of the same nature of the carriers in both types of the strips. The effective interferometer area is, however, defined by the screening of the edge potential in the compressible regions~\cite{fabryperot,litvin1}.

The situation is more intriguing in the fractional QH regime. To preserve the phase coherence, the interference paths should be defined by the states of the same type, which is not obvious at the fractional QH edge. Compressible regions separate incompressible QH states, which are at different local fractional filling factors~\cite{Beenakker,chamon,chklovskyCF,kouwen,cunning}. Each fractional incompressible strip is described by its own ground state, elementary excitations,  and the collective edge excitation modes~\cite{chamon,chklovskyCF,macdonald,wen}. On the other hand, the compressible regions are constructed from "normal" electrons~\cite{Beenakker,chamon}. Thus, it is still an open question, how the compressible/incompressible strips structure affects the phase coherence at the fractional QH edge. 

Here, we experimentally investigate interference effects in transport across a single incompressible strip at the edge of the quantum Hall system by using a Fabry-Perot type interferometer~\cite{fabryperot}. The applied experimental geometry allows us to independently demonstrate the presence of the incompressible strip at the sample edge and to study interference effects in transport across it. We find the interference oscillations in transport across the incompressible strips with local filling factors  $\nu_c=1, 4/3, 2/3$ even at high imbalances, exceeding the spectral gaps. In contrast, there is no  sign of the interference in transport across the principal Laughlin $\nu_c=1/3$ incompressible strip. This indicates, that even at fractional $\nu_c$, the interference effects are caused by "normal" electrons, supporting the earlier prediction~\cite{Beenakker}.  The oscillation's period is determined by the effective interferometer area, which is sensitive to the filling factors because of screening effects.

\section{Samples and technique}

The goal of this experiment is to extend the interference investigations in co-propagating environments~\cite{fabryperot} into the fractional QH regime. For this reason, we concentrate here on the most important features. The details of the interferometer scheme~\cite{fabryperot} can be found in the Appendix section. The well-established methods of the investigations in the quasi-Corbino geometry~\cite{alida,relax} are also described there. 

\begin{figure}
\includegraphics*[width=0.75\columnwidth]{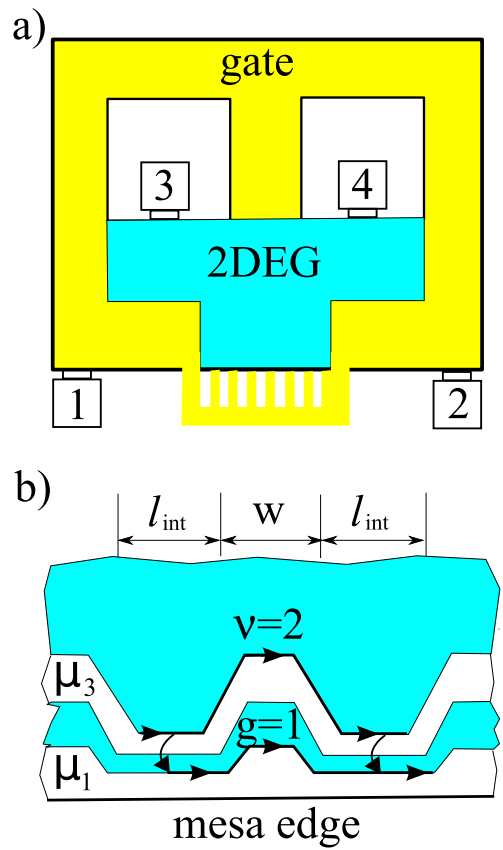}%
\caption{(Color online) (a) Schematic diagram of the sample (not in the scale). The etched mesa edges are shown by thick solid lines. The outer sample dimension is about 2x2~mm$^2$. The inner etched regions (white) are approximately 0.5x0.5~mm$^2$. Light yellow (light gray) areas indicate the split-gate, that covers 2DEG around the inner etched regions and forms a 10~$\mu$m width gate-gap region at the outer mesa edge. Light green (gray) area indicates uncovered 2DEG. The gate-gap region contains a side-gate finger structure, connected to the main gate. The lithographic width of each finger is $w=400$~nm, they are separated by the $l_{int}=400$~nm-wide interaction regions.  Ohmic contacts are denoted by bars with numbers. (b) Schematic diagram of the electron liquid near a side-gate finger (see also Appendix). Light green (gray) areas are the incompressible regions at filling factors $\nu=2$ (in the bulk) and $\nu_c=g=1$ (the incompressible strip at the mesa edge). The latter is wider around a finger region.
Compressible regions (white) are at the electrochemical potentials
of the corresponding ohmic contacts, denoted by bars with numbers in part (a).
Bold lines with arrows indicate two possible paths for an electron around the finger region. The region within these paths defines the effective interferometer area.
\label{sample}}
\end{figure}

Our samples are fabricated from a molecular beam epitaxially-grown GaAs/AlGaAs heterostructure. It contains a 2DEG located 200~nm below the surface. The 2DEG mobility at 4K is  $5.5 \cdot 10^{6}  $cm$^{2}$/Vs  and the carrier density is   $1.63 \cdot 10^{11}  $cm$^{-2}$. Samples are patterned in the quasi-Corbino sample geometry~\cite{alida} with additional gate fingers structure in the gate-gap region~\cite{fabryperot}, see Fig.~\ref{sample} (a).  

In a quantizing magnetic field, at the bulk filling factor $\nu$, the compressible/incompressible strip structure is arising at the mesa edges~\cite{shklovsky}, see Fig~\ref{sample} (b). The quasi-Corbino geometry provides direct investigations of the transport across a single incompressible strip with local filling factor $\nu_c<\nu$, which is determined by the  split-gate, at high imbalances across the strip (see Appendix). Gate fingers structure splits the interaction region into a number of smaller ones. An electron can be transferred across the strip $\nu_c$ at any region with some probability, or can go to the next region along the sample edge. If the structure dimension is smaller than the coherence length $l_c$, the interference between the different paths is possible.

A Fabry-Perot type interferometer~\cite{fabryperot} has serious advantages for fractional QH studies. First of all, it allows investigations in co-propagating environments for the interference paths~\cite{fabryperot}. Also,  several interaction regions  allow to increase the visibility. To obtain the interference, the width of the interaction region $l_{int}$ should be significantly smaller than the  coherence length $l_c$ and the equilibration length~\cite{mueller} $l_{eq}$. They both  are of the same order~\cite{kouwen,cunning,fracdens,goldman,roulleau}      $\sim 10\mu$m in the fractional QH regime, restricting $l_{int}$ to extremely low values.

\begin{figure}
\includegraphics*[width=\columnwidth]{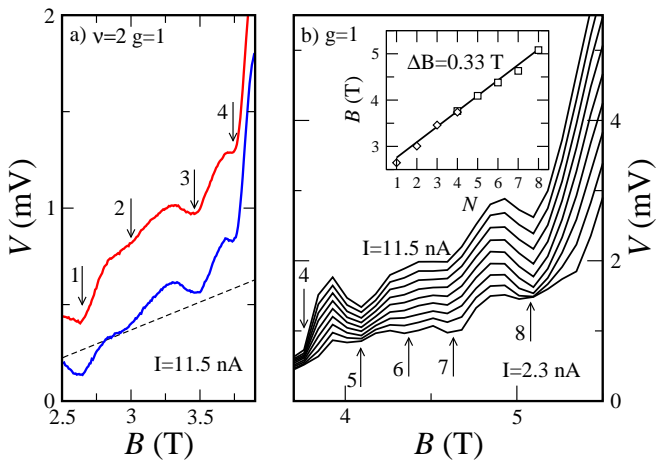}%
\caption{(Color online)  Oscillating behavior for the  transport across $\nu_c=g=1$ integer incompressible strip. (a) Oscillations within the $\nu=2$ QH plateau (solid) for two contact combinations (colors correspond to ones in Fig.~\protect\ref{IV21}). The measurement current equals $I=11.5$~nA. Dash indicates exchange-enhanced Zeeman splitting change in this field range. (b) Oscillations above the $\nu=2$ QH plateau for    different currents ($I$ changes from 2.3~nA to 11.5~nA with 1.15~nA step). The contact combination is of minimal resistance. Inset:  positions of the oscillations as a function of their indexes (open symbols). Solid line indicates a linear  fit with  common period $\Delta B=0.33$~T. \label{oscill21}}
\end{figure}

Our investigations are based on the $I-V$ measurements in the quasi-Corbino geometry~\cite{alida,fabryperot} (see Appendix). There are two ways to study interference effects in transport across a single incompressible strip: 

(i) It is possible to directly reproduce the method of Ref.~\onlinecite{fabryperot}. In this method, the current across the $\nu_c$ incompressible strip is fixed. The voltage drop is traced, while changing the magnetic field withing the bulk QH plateau. Keeping the side-gate voltage constant allows to avoid the resistance fluctuations in charging the sample edge. This method  is very accurate for small oscillations within the bulk QH plateau. It demands  a wide and well-developed QH plateau under the gate, which should be verified by the magnetocapacitance investigations. It cannot be used beyond the bulk QH plateau.

(ii) It is also possible to trace $I-V$ curves at different magnetic fields, tuning the gate voltage to the center of the $\nu_c$ QH plateau.   $V(B)$ dependence can thus be recalculated in this case at any current value. Tuning of the gate voltage affects the potential profile in the finger regions, which is diminishing the accuracy. This method is appropriate either beyond the bulk QH plateau for  strong interference oscillations, or for the weak $\nu_c$ filling factors with narrow QH plateaux (like $4/3$ in our experiment). 

Side-gate fingers are  more appropriate for the fractional QH studies than the top-gate ones, because they do not affect the 2DEG quality in the finger regions. Standard two-point magnetoresistance is used to obtain the electron concentration in the ungated area and to test the contact behavior (below 100~$\Omega$ at low temperature). Magnetocapacitance measurements are performed to determine the available integer and fractional filling factors $\nu_c=g$ under the gate. Two different Ohmic contact combinations~\cite{alida} are tested to verify the results.   There are no interference oscillations for the reference samples without the gate finger structure. The results are independent from the cooling cycle.  The measurements are performed in a dilution refrigerator with a base temperature of 30~mK, equipped with a superconducting solenoid.

\section{Experimental results: integer QH regime}

First of all, we reproduce the results of Ref.~\onlinecite{fabryperot} at integer fillings for the current side-gate fingers geometry, and extend them outside the bulk QH plateau. The consistency of two  investigation  methods, described above, is also tested in this regime.

The result of the interference is presented in Fig.~\ref{oscill21} (a) for integer filling factors $\nu=2, g=1$ for two different contact combinations. The curves are obtained by sweeping the magnetic field at the constant current. They demonstrate clear visible oscillations, equally spaced in the magnetic field, similar to Ref.~\onlinecite{fabryperot}. No such oscillations can be seen for the reference samples without gate finger structure, where the signal follows the exchange-enhanced Zeeman splitting dependence~\cite{fracbutt} (dash in the figure). 

Fig.~\ref{oscill21} (b) demonstrates oscillations in transport across the $\nu_c=g=1$ incompressible strip outside the bulk $\nu=2$ QH plateau.  $V(B)$ dependencies are obtained by recalculating from the $I-V$ curves, which are taken with 83.5~mT step in magnetic field.  The main behavior of the $V(B)$ dependencies differs from one within the $\nu=2$ QH plateau. $V(B)$ jumps up at the plateau edge and is rising non-linear at higher fields.  The oscillation picture strongly depends on the measurement current: the minimum with the index 6 can only be seen at the lowest current, while the ones labeled as 5 and 8 are diminishing at low currents. 

The inset to Fig~\ref{oscill21} (b) shows the positions of the oscillations as a function of their index. The data are obtained  within $\nu=2$ QH plateau (diamonds, from the part (a)) and above it (squares, from the main part (b) ). Experimental points can well be fitted by a single straight line, which corresponds  to $\Delta B=0.33$~T for the period of the oscillations. It indicates that both experimental methods give consistent results for the interference in transport across the $\nu_c=1$ incompressible strip.

\begin{figure}
\includegraphics*[width=0.7\columnwidth]{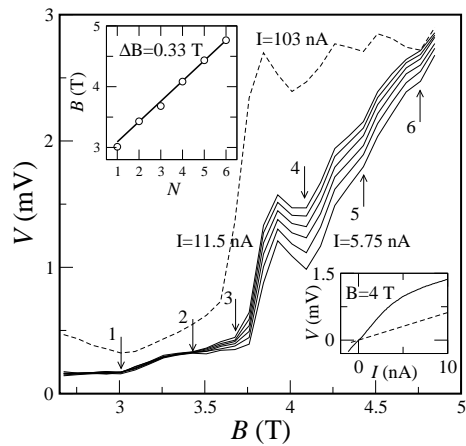}%
\caption{ Main figure: oscillating behavior for transport across $\nu_c=g=4/3$ fractional incompressible strip for different currents $I$. $I$ changes from 5.75~nA to 11.5~nA with 1.15~nA step. Dash indicates the highest current $I=103$~nA. The contact combination is of minimal resistance. Left inset:  positions of the oscillations as a function of  their index (open circles). Solid line fits experimental data with $\Delta B=0.33$~T period. Right inset: an example of the $I-V$ curve (solid). It is typical for transport through the fractional incompressible strip at high imbalances~\protect\cite{fracdens}. $I-V$ curve is non-linear in the whole current range and goes above the equilibrium line (dash).    \label{oscill243}}
\end{figure}

\section{Experimental results: fractional QH regime}

Fig.~\ref{oscill243} shows the oscillations in transport across the fractional $\nu_c=4/3$ incompressible strip in the same field range. Experimental curves are qualitatively different from the integer $\nu_c=1$ case, which seems to be a result of  the qualitatively different $I-V$ curve. The experimental $I-V$ curve is typical for the transport through the fractional incompressible strip at high imbalances~\protect\cite{fracdens},  as demonstrated in the right inset to Fig.~\ref{oscill243}. It is non-linear in the whole current range and goes above the equilibrium line (dash). 

\begin{figure}
\includegraphics*[width=\columnwidth]{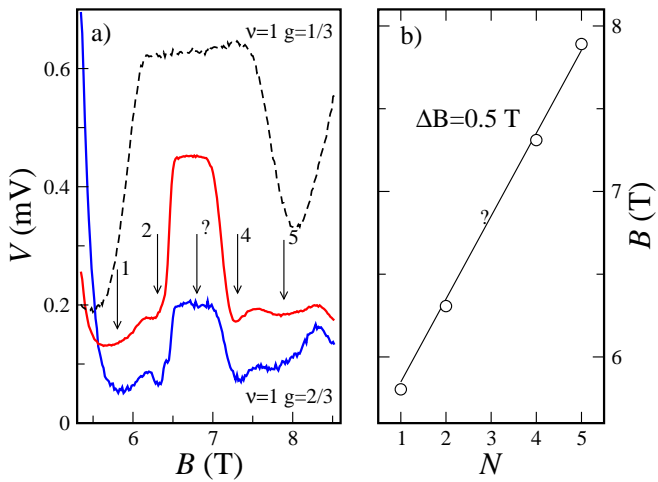}%
\caption{ (a) Oscillating behavior for transport across $g=2/3$  incompressible strip within the $\nu=1$ QH plateau (solid) for two contact combinations (colors correspond to ones in Fig.~\protect\ref{IV21}). The signal for the high-$R$ curve is divided over 3. Dash demonstrates that there are no oscillations for $g=1/3$ (for the sake of the simplicity, only one curve is shown).  The measurement current equals  $I=11.5$~nA. The $\nu=1$ bulk QH plateau ranges between 5.5~T and 8.2~T.  (b) Positions of the oscillations as a function of their index (open circles). Solid line fits experimental data with $\Delta B=0.5$~T period.
 \label{oscill123}}
\end{figure}

$V(B)$ dependencies are recalculated from the $I-V$ curves, which are taken with  83.5~mT step in magnetic field. Similarly to the integer $\nu_c=1$ case, $V(B)$ behavior is different within $\nu=2$ plateau and above it. $V(B)$  jumps up at the plateau edge, and the oscillations are much more pronounced above $\nu=2$ QH plateau. Experimental curves in this region are very sensitive to the current, but the oscillations are clearly visible at any current. Above the $\nu=2$ QH plateau, they even survive at extremely high currents (dash). The oscillation's period $\Delta B=0.33$~T  is equal to one for the integer $\nu_c=1$ case, see the left inset to Fig.~\ref{oscill243}.

The data in Fig~\ref{oscill123} (a) correspond to the transport across $\nu_c=2/3$ (solid) or $\nu_c=1/3$ (dash) fractional incompressible strips. They are obtained by sweeping the magnetic field  within the $\nu=1$ integer QH plateau, at constant current and  gate voltage. The data for $\nu_c=1/3, 2/3$ are  in the same magnetic field range and at the same configuration of the compressible/incompressible strips in the gate-gap region, because of the same $\nu=1$ in the bulk. Thus it is not surprising, that the curves  are qualitatively similar: they are                non-monotonic, and more or less symmetric in respect to the bulk  $\nu=1$ plateau center ($B=6.74$~T).

There is still a strong difference between experimental curves for $\nu_c=2/3$ and $\nu_c=1/3$. Two curves for $\nu_c=2/3$ (solid, two different contact combinations) demonstrate oscillations while moving away from the $\nu=1$ QH plateau center, see Fig.~\ref{oscill123} (a). In contrast, there are no oscillations in transport across the $\nu_c=1/3$ fractional incompressible strip (dash). The experimental curve for $\nu_c=1/3$ is flat within the bulk $\nu=1$ QH plateau and goes down at the plateau edges.

\begin{figure}
\includegraphics*[width=0.75\columnwidth]{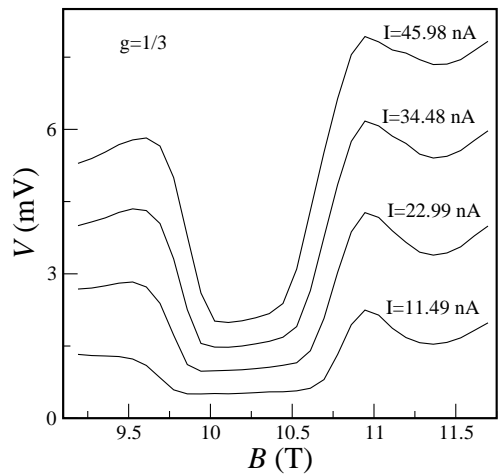}%
\caption{ Transport across $g=1/3$  incompressible strip around the $\nu=2/3$ bulk QH plateau (solid). No oscillations can be seen in the curves. The $\nu=2/3$ bulk QH plateau ranges between 9.5~T and 10.7~T.   \label{oscill2313}}
\end{figure}

We check, that there are no oscillations in transport across  $\nu_c=1/3$ at higher fields (up to 11.7~T), around the bulk  $\nu=2/3$  QH plateau, see Fig.~\ref{oscill2313}. $V(B)$ dependencies are recalculated from the $I-V$ curves, which are taken with 83.5~mT step in magnetic field. The experimental $V(B)$ curve is flat within the bulk $\nu_c=2/3$ QH plateau and is rising up at the plateau edges. Even if we attribute two deep minima at the $\nu=1$ QH plateau edges in Fig.~\ref{oscill123} (a) to the interference effects, there are no corresponding features in Fig.~\ref{oscill2313} neither in period, nor in the magnetic field positions. This behavior is completely contrary to one for $\nu_c=1$, see Fig.~\ref{oscill21}, which indicates, that we do not see interference effects at $\nu_c=1/3$. On the other hand,  $V(B)$  jump at the bulk plateau edge is not surprising (see Figs.~\ref{oscill21},\ref{oscill243}) and seems to be determined by the screening effects~\cite{shklovsky}.

The oscillations at $\nu_c=2/3$ are situated to both sides from the center of the $\nu=1$ QH plateau. It is difficult to say anything about the center itself. There is some sign (marked by "?") at the curve that belongs to the minimal resistance contact combination, see Fig.~\ref{oscill123} (a). There is, however, nothing for the other curve. Other oscillations are similar for both curves. Fig.~\ref{oscill123} (b)  demonstrates a linear fit of the oscillation positions (circles). It gives the oscillation's period to be equal to $\Delta B=0.50$~T. It is supposed in the figure, that there is a missing oscillation at the center of the $\nu=1$ plateau, which is strongly supported by the high quality of the linear fit. In the opposite case the period would be even higher than $0.50$~T, so it is clearly different from the $\nu_c=1,4/3$ cases.

We have no data for the transport across $\nu_c=2/3$ above the bulk $\nu=1$ plateau, because in the present samples the bulk $\nu=1$ QH plateau is very close to the next bulk $\nu=2/3$  QH plateau. It makes to be impossible to study transport across $\nu_c=2/3$ above $\nu=1$.

\section{Discussion}

We can summarize our experimental results: (i) while sweeping the magnetic field, we see interference oscillations in transport through the $\nu_c=1$ integer and $\nu_c=4/3,2/3$ fractional incompressible strips. No oscillations can be seen in transport across $\nu_c=1/3$. (ii) The observed oscillations are periodic in the magnetic field. The period is the same for transport across $\nu_c=1, 4/3$ and equals to $\Delta B=0.33$~T. The period is greater in 1.5 times for transport across the $\nu_c=2/3$ incompressible strip.  (iii)  The interference oscillations can be observed at high imbalances.  The applied imbalances exceed  the spectral gap in  the $\nu_c$ incompressible strip.

It seems to be quite natural to formulate, that we observe the interference of quasiparticles with fractional charge $e^*$. The oscillation's period is given~\cite{goldman} by $\Delta B = h/e^* S$, where $S$ is the interferometer area.   The difference in the oscillation's periods can thus be attributed to different quasiparticle charges $e^*$ for the filling factors $\nu_c=1, 2/3, 4/3$.
However, a deeper analysis is important, because the oscillations are present only at filling factors $\nu=2/3,4/3$ not from the principal Laughlin sequence.  Also, the effective interferometer area $S$ can not be regarded as  constant in our experiment.

\begin{figure}
\includegraphics*[width=\columnwidth]{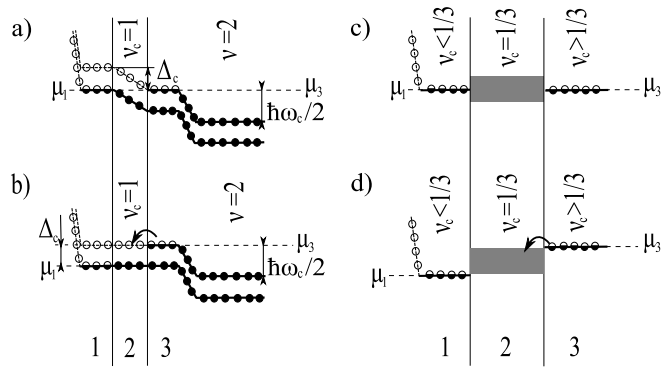}%
\caption{Schematic diagram of  the  energy levels in the active sample area.  Filled circles represent the fully occupied electron states in the incompressible strip (2) and in the bulk. Half-filled circles indicate the partially occupied electron states in the compressible strips (1 and 3).  Open circles are for the empty states. $\Delta_c$ is the potential jump in the $\nu_c$ incompressible strip. Pinning of the Landau sublevels to the Fermi level (shot-dash) is shown in the compressible regions at electrochemical potentials $\mu_1$ and $\mu_3$. (a) Integer filling factors $\nu=2, g=1$. Equilibrium situation $\mu_3=\mu_1$, no electrochemical imbalance is applied across the incompressible strip. (b) Integer filling factors $\nu=2, g=1$. Flat-band conditions for the electrochemical imbalance $\mu_3-\mu_1=\Delta_c$. Arrow indicates a new way for an electron  along the energy level, without spin flip. (c) Fractional filling factor $\nu_c=1/3$. Equilibrium situation $\mu_3=\mu_1$, no electrochemical imbalance is applied across the incompressible strip. A single-particle picture of electron levels is not applicable within the fractional incompressible strip, which is indicated by the gray region.  (d) Fractional filling factor $\nu_c=1/3$. High electrochemical imbalance is applied across the incompressible strip. Arrow indicates adding an electron to the fractional state.  
\label{levels}}
\end{figure}

Let us start from the simplest case of integer $\nu_c=1$. The interference effects can only be seen at high imbalances across the $\nu_c=1$ incompressible strip~\cite{fabryperot}, in contrast to     the experiments~\cite{heiblum,goldman,litvin,glattli}. At high electrochemical imbalance, which  is about the spectrum (Zeeman) gap $\Delta_c$, the flat-band situation is realized at the sample edge~\cite{alida,relax}, see Fig.~\ref{levels} (a,b). Electrons can be transferred across the incompressible strip by  tunneling along the same energy level~\cite{barrier}. The electron's spin flip is not necessary in this case, so the transfer process is partially coherent in two neighbor interaction regions. The interference paths are formed by transport across the edge~\cite{alida,relax,fracdens} in the interaction regions and by the transport along   the outer borders of the incompressible strips~\cite{tauless,gerhards} between the interaction regions, see  Fig.~\ref{sample} (b) and Appendix.

The situation is more complicated in the fractional QH regime.  Every  fractional incompressible strip  is characterized by the many-body ground state, elementary excitations,  and edge collective excitation modes~\cite{laughlin,haldane,obzor,chamon,chklovskyCF,macdonald,wen}. The incompressible strip separates two compressible ones, which are constructed from "normal" electrons at the Fermi level, with the same spin and Landau indexes, see Fig.~\ref{levels}, (c).  In contrast to the integer situation, the charge transfer across the fractional incompressible strip can not be described in a single-particle picture. Adding an electron to the fractional edge leads to the creation of several fractional elementary excitations within the strip.  The edges of the fractional strip are characterized by collective  modes~\cite{chamon,wen}, which are excited if an electron is added to the fractional edge or removed from it. These collective modes define the tunnel density of states~\cite{wen,chang,grayson,fracdens}, which is reflected in a power-law $I-V$ curve~\cite{chang,grayson,fracdens,fraceq} (see also the right inset to Fig.~\ref{oscill243}). 

We do not see interference effects in transport across the incompressible strip at principal Laughlin 1/3 filling factor. On  the other hand, there are clear visible oscillations in transport across  2/3 and 4/3 strips. The fractional QH states 2/3 and 4/3 are very similar in Laughlin's~\cite{laughlin,haldane,obzor} theory. In contrast to the principal 1/3 state, they are constructed as a quasi-hole (or, correspondingly, quasi-electron) ground state at \textit{filled} first Landau level. The ground state structure is reflected in the excitation spectrum~\cite{macdonald}. The edge gapless modes also follow for the ground state structure of the fractional strip~\cite{wen,chamon,macdonald}. There is only one excitation branch for the principal Laughlin 1/3 filling factor. However, there are two excitation branches for the non-principal states like 2/3 and 4/3, because of the ground state structure. It is important, that in these cases one excitation branch belongs to electrons from the filled Landau level~\cite{macdonald,wen}.

We can summarize, that the interference oscillations are observed in transport between two compressible strips, constructed from \textit{electrons}. It is an \textit{electron}, which is transferred across the incompressible strip even for the fractional $\nu_c$. The phase coherence is provided by the \textit{electron} excitation modes in this case.

The obtained period of the oscillations $\Delta B$ is only determined by the interferometer  effective area $S$. It can vary for different filling factors, because of screening at the sample edge. The effective area $S$ is defined by the effective width $w$ of the finger region and by the corresponding length $L$ across the edge in Fig.~\ref{sample} (b). The latter is the sum of the compressible and incompressible strips widths in the finger region.  Both $w$ and $L$ are sensitive to the screening effects at the sample edge. They are dependent~\cite{shklovsky} not only on the $\nu_c$ filling factor, but also on the filling factor in the bulk $\nu$. It may be a reason to obtain the same oscillation's period $\Delta B=0.33$~T for the $\nu_c=1$ and $\nu_c=4/3$, because of the same bulk filling. The  oscillation's period $\Delta B=0.33$~T allows to estimate the effective area to $S \sim 10^{-2} \mu\mbox{m}^2$. The oscillation's period is in 1.5 times higher for $\nu=1, g=2/3$, that corresponds to the 1.5 smaller effective area for $\nu_c=2/3$. This is, possibly, because of the smaller $\nu_c=2/3$ and $\nu=1$.

The above considerations support the earlier prediction~\cite{Beenakker}, that the interference experiment at the smooth sample edge is not sensitive to the fractional charge. They can not be directly applied to other types of interferometers~\cite{heiblum,goldman,litvin,glattli}, because of different experimental geometry.  We believe, however, that the presence of the compressible regions and screening effects should be important in these experiments also.

\section{Conclusion}

As a conclusion, we experimentally investigated interference effects in transport across a single incompressible strip at the edge of the quantum Hall system by using a Fabry-Perot type interferometer~\cite{fabryperot}. The applied experimental geometry allows us to independently demonstrate the presence of the incompressible strip at the sample edge and to study interference effects in transport across it. We found the interference oscillations in transport across the incompressible strips with local filling factors  $\nu_c=1, 4/3, 2/3$ even at high imbalances, exceeding the spectral gaps. In contrast, there was no  sign of the interference in transport across the principal Laughlin $\nu_c=1/3$ incompressible strip. This is a strong evidence, that even at fractional $\nu_c$, the interference effects are caused by "normal" electrons, supporting the earlier prediction~\cite{Beenakker}.  The oscillation's period is determined by the effective interferometer area, which is sensitive to the filling factors because of screening effects.

\acknowledgments

We wish to thank  V.T.~Dolgopolov for fruitful discussions. We gratefully
acknowledge financial support by the RFBR, RAS, the Programme "The
State Support of Leading Scientific Schools". E.V.D. is also supported by the President's grant MK-1678.2008.2.

\section{Appendix. Experimental details}

Samples are patterned in the quasi-Corbino sample geometry~\cite{alida} with additional gate fingers structure in the gate-gap region~\cite{fabryperot}, see Fig.~\ref{sample}, (a). Each sample has two macroscopic ($\approx 0.5\times 0.5\mbox{mm}^2$) etched regions inside. Ohmic contacts are made to both the inner mesa edges and to the outer one.   A split-gate, encircling the etched regions, is used to connect these independent mesa edges in a controllable way. The gate-gap region at the outer mesa edge is of microscopic size  (10~$\mu$m in the present samples).  Standard two-point magnetoresistance is used to obtain the electron concentration in the ungated area and to test the contact behavior (below 100~$\Omega$ at low temperature). Magnetocapacitance measurements are used to determine the available integer and fractional filling factors $g$ under the gate.

In a quantizing magnetic field, at the bulk filling factor $\nu$, the compressible/incompressible strip structure is arising at the mesa edges~\cite{shklovsky}, see Fig~\ref{sample}, (b). By depleting 2DEG under the gate to a lower integer or fractional filling factor $g<\nu$, some of the strips (with local filling factors $\nu_c>g$) are redirected to another mesa edge. The others with $\nu_c<g$ are still at the etched mesa edges. A single incompressible strip with $\nu_c=g$ is directly connected to  the incompressible QH state under the gate.  As a result, it separates the compressible strips, originating from inner and outer mesa edges in the gate-gap region.   These compressible strips are at the electrochemical potentials of the corresponding Ohmic contacts~\cite{shklovsky}. By applying  dc bias to them, we directly apply it across the incompressible strip with $\nu_c=g$. As a result, the quasi-Corbino sample geometry~\cite{alida} allows to directly detect the incompressible strip with the local filling factor $\nu_c$ at the sample edge and to study transport across it.

A side-gate finger structure  is made along the mesa edge in the gate-gap region, see Fig.~\ref{sample} (\textit{cp.} Ref.~\onlinecite{fabryperot}). The structure is formed by 10 fingers of the width $w=0.4 \mu\mbox{m}$, separated by $0.5 \mu\mbox{m}$ intervals.  Side-gate fingers are connected to the main gate, so they are at the negative gate potential. It moves 2DEG away from the edge in the finger region and thus increases the depletion region at the edge, making the edge profile \textit{smoother}. In other words, the negative finger potential repels  the strips away from the edge  and makes them \textit{wider}, see Fig.~\ref{sample}, (b). Thus, the side-gate finger structure modulates the transport probability  across the $\nu_c=g$ incompressible strip. Side-gate fingers are  more appropriate for the fractional QH studies than the top-gate ones~\cite{fabryperot}, because they do not affect the 2DEG quality in the finger regions.

\begin{figure}
\includegraphics[width=0.8\columnwidth]{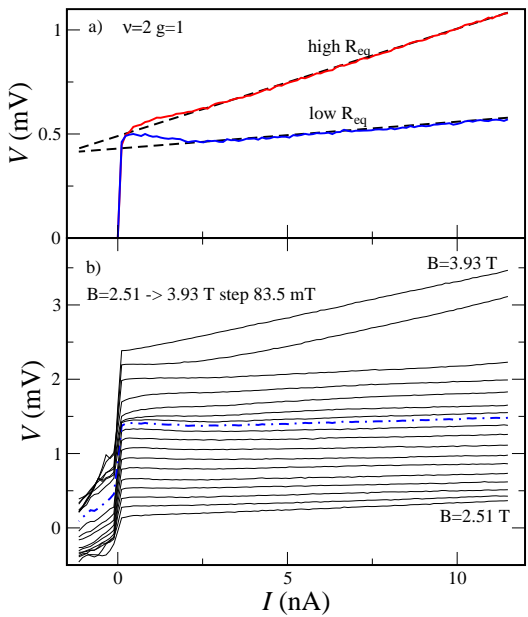}%
\caption{(Color online) $I-V$ curves (solid) for filling factors $\nu=2, g=1$. (a) $I-V$ traces at the center of $\nu=2$ QH plateau ($B=3.34$~T) for two different contact combinations (red for the high $R_{eq}=2 h/e^2$ and blue for the low $R_{eq}=0.5 h/e^2$ contact combinations). The linear fit of the positive branches gives equilibrium slopes $R=2 h/e^2, 0.5 h/e^2$ but slightly different threshold voltages $V_{th}=0.49$~mV and $V_{th}=0.43$~mV. The onset voltage is equal for both curves. (b) $I-V$ evolution while changing magnetic field from 2.51~T to 3.93~T with 83.5~mT step. The contact combination correspond to the minimum equilibrium resistance. Curves are shifted vertically on 0.1~mV for clarity. Blue dash-dot is the curve at the center of the plateau, depicted in (a) part (blue solid).  \label{IV21}}
\end{figure}

The dissipativeless (diamagnetic) current, flowing along the sample edge, is carried in the incompressible regions only, because the group velocity is zero in compressible regions~\cite{tauless,gerhards}. Out of balance, however, the border position between the compressible and incompressible strip is changed. It is the electrons in this region that define the transport current along the sample edge~\cite{tauless,gerhards}.  The transport \textit{across} the sample edge is carried by tunneling through the incompressible strip and by diffusion within the compressible strip. For our experimental geometry, an electron can more likely be transferred across the incompressible strip   between the two finger regions, or can go to the next interaction region~\cite{suppose}, see Fig.~\ref{sample}, (b). For a particular electron it is impossible to predict the path.  If the finger region width $w$ is smaller than the coherence length $l_c$, the transport current is determined by the sum of the path's probabilities. The phase shift between the paths  can be controlled by the magnetic field sweep~\cite{fabryperot}. 

We study $I-V$ curves in four-point contact scheme, by applying a \textit{dc} current between one pair of inner and outer contacts (1-3 or 2-4 in Fig.~\ref{sample}, (a)), and measuring the \textit{dc} voltage between another pair of inner and outer contacts (2-4 or 1-3, correspondingly). These two contact combinations are characterized by the minimum (the former combination) or by the maximum (the other one)  equilibrium resistances~\cite{buttiker,alida}.

The experimental $I-V$ curves are shown in  Fig.~\ref{IV21} for integer filling factors $\nu=2, g=1$. Each curve is strongly non-linear and asymmetric, as it is usual for the transport across the integer incompressible strip~\cite{alida}.  The positive branch starts from the onset voltage $V_{on}\sim 0.5$~mV. It is linear above $V_{on}$ with equilibrium slope $R=0.5 h/e^2 \mbox{ or } 2 h/e^2$,  see Fig.~\ref{IV21}, (a). $R$ is defined by the particular contact combination and can be easily calculated~\cite{alida}. One feature is specific for samples with the gate finger structure in the gate-gap: extrapolation of the positive branch to zero current gives the threshold voltage $V_{th}$, which differs from  the onset voltage $V_{on}$. The reason is obvious: onset voltage $V_{on}$ is defined by the potential jump in the $\nu_c$ incompressible strip only, and is the same for different contact combinations, see Fig.~\ref{IV21}, (a). It follows the Zeeman splinning while increasing the magnetic field. $V_{th}$, determined from the extrapolation, 
reflects the threshold for the \textit{flowing} current. It is sensitive to the interference conditions~\cite{fabryperot}, so $V_{th}$ oscillates around $V_{on}$ while sweepping the magnetic field.

In Fig.~\ref{IV21} (b) the evolution of the $I-V$ curve is shown while moving along $\nu=2$ QH plateau and keeping $g=1$ under the gate. The slope of the positive branch $R$ is constant around the center of the $\nu=2$ plateau, while it is higher at the plateau edges.   It can be seen from Fig.~\ref{IV21} (b), that $R$ is rising for high currents first, that leads to the 'kink' on the $I-V$s. Thus, we can expect the results  to be independent from the particular current value within the QH plateau, while there should be some dependence at the plateau edges and beyond them. For this reason, the methods of the  interference investigations should be different within the QH plateau and beyond it, as described in the main text.


\begin{thebibliography}{99}
\bibitem{barrier} It was shown in Ref.~\protect\onlinecite{fracdens} that the transition probability is always below 1 even at imbalances exceeding the spectrum gap.
\bibitem{suppose} We suppose here that the transfer probability is \textit{much} lower in the finger region, whis is a good approximation  because of the exponential dependence of the probability on the strip's width.
\end{thebibliography}
\end{document}